\documentclass[twocolumn]{aastex701}

\newcommand\latex{La\TeX}

\accepted{July 11, 2025}
\usepackage{comment}
\usepackage{soul}
\usepackage{graphicx}   % for including graphics
\usepackage{float}

\begin{document}
\title{Puzzling Variation of Gamma Rays from the Sun over the Solar Cycle Revealed with \textit{Fermi}-LAT}

\author[0000-0002-2028-9230]{A.~Acharyya}
\affiliation{Center for Cosmology and Particle Physics Phenomenology, University of Southern Denmark, Campusvej 55, DK-5230 Odense M, Denmark}
\email{atreya@cp3.sdu.dk}
\author[]{A.~Adelfio}
\affiliation{Istituto Nazionale di Fisica Nucleare, Sezione di Perugia, I-06123 Perugia, Italy}
\email{Andrea.Adelfio@pg.infn.it}
\author[0000-0002-6584-1703]{M.~Ajello}
\affiliation{Department of Physics and Astronomy, Clemson University, Kinard Lab of Physics, Clemson, SC 29634-0978, USA}
\email{majello@clemson.edu}
\author[0000-0002-9785-7726]{L.~Baldini}
\affiliation{Universit\`a di Pisa, Dipartimento di Fisica E. Fermi, I-56127 Pisa, Italy}
\affiliation{Istituto Nazionale di Fisica Nucleare, Sezione di Pisa, I-56127 Pisa, Italy}
\email{luca.baldini@pi.infn.it}
\author[0000-0001-7233-9546]{C.~Bartolini}
\affiliation{Istituto Nazionale di Fisica Nucleare, Sezione di Bari, I-70126 Bari, Italy}
\affiliation{Universit\`a degli studi di Trento, via Calepina 14, 38122 Trento, Italy}
\email{chiara.bartolini@ba.infn.it}
\author[0000-0002-6954-8862]{D.~Bastieri}
\affiliation{Istituto Nazionale di Fisica Nucleare, Sezione di Padova, I-35131 Padova, Italy}
\affiliation{Dipartimento di Fisica e Astronomia ``G. Galilei'', Universit\`a di Padova, Via F. Marzolo, 8, I-35131 Padova, Italy}
\affiliation{Center for Space Studies and Activities ``G. Colombo", University of Padova, Via Venezia 15, I-35131 Padova, Italy}
\email{denis.bastieri@pd.infn.it}

\author[0000-0002-6729-9022]{J.~Becerra~Gonzalez}
\affiliation{Instituto de Astrof\'isica de Canarias and Universidad de La Laguna, Dpto. Astrof\'isica, 38200 La Laguna, Tenerife, Spain}
\email{jbecerragonzalez@gmail.com}
\author[0000-0002-2469-7063]{R.~Bellazzini}
\affiliation{Istituto Nazionale di Fisica Nucleare, Sezione di Pisa, I-56127 Pisa, Italy}
\email{ronaldo.bellazzini@pi.infn.it}

\author[]{B.~Berenji}
\affiliation{California State University, Los Angeles, Department of Physics and Astronomy, Los Angeles, CA 90032, USA}
\email{bijan.berenji@fizisim.com}
\author[0000-0001-9935-8106]{E.~Bissaldi}
\affiliation{Dipartimento di Fisica ``M. Merlin" dell'Universit\`a e del Politecnico di Bari, via Amendola 173, I-70126 Bari, Italy}
\affiliation{Istituto Nazionale di Fisica Nucleare, Sezione di Bari, I-70126 Bari, Italy}
\email{elisabetta.bissaldi@ba.infn.it}
% ...

\begin{abstract}
The steady-state gamma-ray emission from the Sun is thought to consist of two emission components
due to interactions with Galactic cosmic rays: (1) a hadronic disk component, and
(2) a leptonic extended component peaking at the solar edge and extending into the heliosphere. The flux of these components is expected to vary with the 11-year solar cycle, being highest during solar minimum
and lowest during solar maximum, as it varies with the cosmic-ray flux. No study has yet
analyzed the flux variation of each component over solar cycles.

In this work, we measure the temporal variations of the flux of each component over 15 years of \textit{Fermi} Large Area Telescope observations and compare them with the sunspot number and Galactic cosmic-ray flux from AMS-02 near Earth.

We find that the flux variation of the disk anticorrelates with the sunspot number and correlates with cosmic-ray protons, as expected, confirming its emission mechanism. In contrast, the extended component exhibits a more complex variation: despite an initial anticorrelation with the sunspot number, we find neither anticorrelation with the sunspot number nor correlation with cosmic-ray electrons over the full 15-year period.
This most likely suggests that cosmic-ray transport and modulation in the inner heliosphere are unexpectedly complex and  may differ for electrons and protons or, alternatively, that there is an additional, unknown component of gamma rays or cosmic rays.

These findings impact space weather research and emphasize the need for close monitoring of Cycle~25 and the ongoing polarity reversal.

\end{abstract}




\section{Introduction} \label{sec:intro}

Over the last fifteen years, the Sun has been observed to be a gamma-ray source in its quiet state \citep[][]{Orlando2008}, i.e., in its non-flaring state or non-flaring regions. 
\citet{Hudson} and \citet{Seckel91} made the first calculations of the gamma-ray emission from pion decay by Galactic Cosmic Ray (CR) cascades in the solar atmosphere. 
This emission was expected to be confined to the region of the solar disk. The flux was expected to vary during the solar cycle, as observed for the lunar gamma-ray flux \citep{Thompson}, which is produced by similar mechanisms. 
The existence of a spatially extended inverse Compton (IC) component from scattering by CR electrons on solar photons was theorized  by \cite{Moskalenko} and \cite{Orlando2006} independently. This broad emission was expected to be brighter at the solar edges and roughly inversely proportional to the angular distance from the Sun, as recently confirmed by refined IC models \citep{stellarics}. 
Recent theoretical studies have also focused on improving models of the disk component \citep[e.g.,][]{Zhou, Nibl,  Mazziotta, Becker, Guti, Hudson20, Li, LiBeacom2024}. 

Due to their association with CRs, the intensity of both the disk and extended components of the gamma-ray emission from the quiet Sun is predicted to vary over the solar cycle and, in particular, to anticorrelate with the sunspot number (SSN), and to correlate with the CR flux as measured near Earth. 

The first evidence of the gamma-ray emission from the quiet Sun was found by \cite{Orlando2008} by analyzing the entire archival EGRET data. However, the limited sensitivity of EGRET precluded the investigation of flux variation over time.
The launch of the \textit{Fermi} Gamma-Ray Space Telescope in 2008 enabled observations of the quiet Sun with high statistical significance. 
Observations of solar emission during the first 18 months of the \textit{Fermi} Large Area Telescope (LAT) mission, a time period during solar minimum, were reported in \cite{Abdo2011} 
and were in agreement with \cite{Orlando2008}. 
Variations of the solar flux with solar activity were first reported by \citet{Ng} with six years of \textit{Fermi}-LAT observations. They found the gamma-ray flux of the solar emission to be larger during solar minimum than during solar maximum. More recent works \citep{Tang,Linden2022} updated the analysis over almost the entire 11-year cycle and found that the solar flux varies by nearly a factor of two between the solar maximum and the solar minimum. 
Moreover, \cite{Linden2022} found that the variation in the 0.1–10 GeV range is not dependent on energy, differing from what was expected.
More recent observations with HAWC \citep{HAWC} showed the solar flux extending to TeV energies during solar minimum, and an upper limit during solar maximum was observed.

In summary, the observed solar gamma-ray flux variations have confirmed a trend of anticorrelation with the solar activity.
However, the extended IC component has not been included in these studies. Hence, contamination of the disk flux measurement by the extended component is not excluded in previous works. To date, no studies have examined the correlation of the two distinct components with CR proton and electron fluxes, or their anticorrelation with solar activity. 
In this letter, we report results for 15 years of \textit{Fermi}-LAT observations of the Sun, allowing us to distinguish the two solar components and to study their flux variations over Solar Cycle 24 and the first years of Solar Cycle 25.

\section{Method} \label{sec:method}
We use observations with \textit{Fermi}-LAT \citep{Atwood}, which detects gamma rays from 20~MeV to over 300~GeV. In survey mode, the LAT observes the full sky approximately every 3 hours with an almost uniform daily exposure. This capability, combined with its large field of view of ~2.4 sr (at 1 GeV) and high sensitivity, allows us to observe the Sun daily.  

\subsection{Data Selection} \label{sec:data}
We analyze 15 years of \textit{Fermi}-LAT observations, specifically from 2008 August 4 to 2023 June 30 in the 70~MeV -- 70~GeV energy range. We select the Pass 8 SOURCE event class \citep{pass8, bruel} and P8R3$\textunderscore$SOURCE$\textunderscore$V6 instrument response functions\footnote{P8R3 data do not contain the residual background events near the Ecliptic present in P8R2, which could have affected the analysis}, and excluded times when the LAT was within the South Atlantic anomaly. To reduce contamination from cosmic-ray interactions with the upper atmosphere, events with zenith angles larger than 90$^\circ$ are excluded.
Because the Sun is moving across the sky, the analysis of its emission requires special handling. 
In addition to the standard \textit{Fermi}-LAT Science Tools package\footnote{Fermitools version 2.2.0, \\ https://fermi.gsfc.nasa.gov/ssc/data/analysis/software/}, dedicated tools are used where data are selected in a  moving frame centered on the instantaneous position of the Sun, which is computed using an interface to the JPL ephemeris libraries\footnote{https://ssd.jpl.nasa.gov/horizons/}.
As in \citet{Abdo2011}, to further reduce contamination while maintaining useful data, 
we apply a carefully customized selection: data are excluded when the Sun is within 30$^\circ$ of the Galactic plane,  20$^{\circ}$ of the Moon, and 10$^{\circ}$ of the four bright sources along its path (3C 454.3, PSR~J1620$-$4927, Geminga, and the Crab Nebula).
In addition, all solar flares listed in the LAT Solar Flare Catalog  \citep{SolarFlareCat} are removed. Also, to remove any contamination from fainter solar flares and other flaring or bright sources close to the path of the Sun, in this work we develop a dedicated procedure involving the monitoring of light curves in nested regions centered on the Sun. After inspection, we remove the time intervals when short variations of light curve fluxes exceed 5$\sigma$ from the average flux value, using 3-hour time bins. 
These cuts produce a very clean event sample by removing about 55$\%$ data in total.

\subsection{Background Evaluation} \label{sec:BKG}
After applying our data cuts, there may still be possible contamination in the region around the Sun from diffuse Galactic and isotropic gamma-ray emissions, as well as from weak point sources along the ecliptic. This contamination is referred to as background. Correct evaluation of the background is essential for analyzing solar emission, as IC emission extends beyond 20$^{\circ}$ from the Sun \citep{Orlando2006}. 

As in \citet{Abdo2011}, we determine the background by applying the off-source (or ``fake'' source) method: an imaginary source follows the path of the Sun along the ecliptic but at a different time, which means at a different distance from the real Sun. 
The data sample we use for the background determination is centered on the off-source position and has the same selection cuts applied to the data sample of the Sun. 
For a careful evaluation of the background, we compute seven fake sources from
90$^{\circ}$ to 270$^{\circ}$ from the Sun in steps of 30$^{\circ}$.
Moreover, we divide the 15-year dataset into 2-year sub-samples, each starting six months after the previous one, on which we perform the binned maximum likelihood technique as described by \citet{Mattox}, using gtlike as included in the Fermitools. We do it by iteratively adjusting the energy spectrum in order to maximize the likelihood of the data given the model.
We find the seven background fluxes to be within $\pm$3\% of their average values. Following \citet{Abdo2011}, this 3\% variation defines the systematic uncertainty on the solar components.  
Then, we use the output of the likelihood analysis for each time interval as the background model for the corresponding  data set centered on the Sun.

\subsection{Analysis of the Solar Components}
We evaluate the gamma-ray flux variation over time by dividing the 15-year dataset into 2-year sub-samples, each each starting six months after the previous one, similar to the procedure for the background.
For the binned maximum likelihood analysis we use a bin size of 0.1$^{\circ}$ and the energy dispersion (edisp\_bins=$-$2).
Three components are considered simultaneously in the analysis: the solar disk component, the solar extended component, and the background. The latter is fixed as previously described and masked over the solar disk.
We model the disk with emission confined within a radius of 0.265$^{\circ}$ from the solar center,  corresponding to the average radius of the Sun. We model the extended component with emission starting from the edge of the solar disk and decreasing inversely with the elongation angle from the Sun, following solar IC emission models \citep{Orlando2006,Orlando2008,stellarics}, as in \cite{Abdo2011}. Both energy spectra are modeled with a log-parabolic function. The solar model components are convolved with the energy-dependent point-spread function, while the background component is not, because it is derived from observational data.

%%%%%%%%%

\section{Temporal Variation Results} \label{Results}
The two components are clearly separately measured\footnote{The significance of the detection of each component is verified for each data point, being the TS$>$300.} and their flux variations are integrated over 2-year sub-samples.
Then, the disk and extended component variations are compared with the flux of CR protons \citep{AMSprotons} and CR electrons \citep{AMSelectrons}, respectively, and with the SSN\footnote{SSN data are downloaded from the World Data Center SILSO,  https://www.sidc.be/SILSO/DATA/SN$\_$ms$\_$tot$\_$V2.0.txt}.

The correlation between these datasets (solar flux, CR flux, SSN) is visually inspected and statistically investigated by calculating the Pearson correlation coefficient \citep{Pearson1920, Pearson1931, Pearson1932} and the z-transformed discrete correlation function (zDCF)\footnote{Because the Pearson coefficient may not be accurate for non-linear relationships \citep{Yule, Scargle, Riechers} we also use zDCF, which is better suited for temporal cross-correlations and commonly applied to non-linear cases.} \citep{Alexander, Edelson, Blandford}. The eventual time lag between datasets is also evaluated. For a proper correlation study, the datasets of \textit{Fermi}-LAT, the daily CR fluxes, and the monthly SSN are interpolated or averaged over the same time bin of 30 days. 
We test the robustness of the results using non-overlapping 2-year sub-samples, each starting six months after the previous one, to verify that the results are not sensitive to the choice of sub-sample start times. Since solar flux variations in sub-samples are integrated over 2-year periods starting every six months, an apparent trend on a timescale shorter than 2 years should not be interpreted as physically meaningful.

The results for the two components are reported below. 

% Full-page figure environment that spans both columns
\begin{figure*}[htbp]
    \centering
    \makebox[\textwidth][c]{
        \begin{minipage}[b]{0.45\textwidth} % 45% of text width
            \centering
            \includegraphics[width=\textwidth]{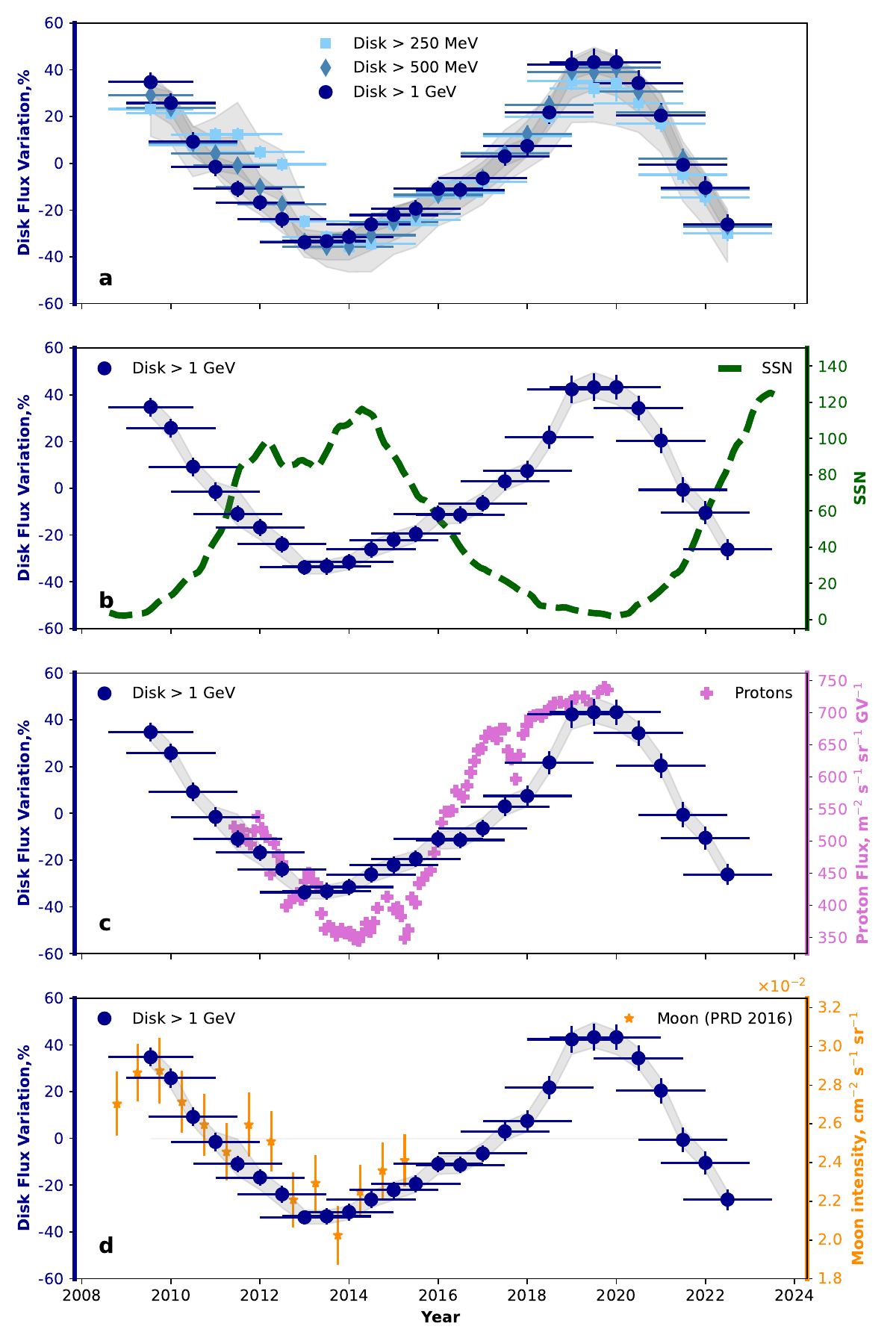} % Replace with your image
            \caption{Panel (a): Percentage of the flux variation of the solar disk component above 25~MeV, 500~MeV, and 1~GeV versus time in year. For normalization, the average flux values used are 9.15, 4.69, 2.23 ($\times$ 10$^{-8}$ cm$^{-2}~$s$^{-1}$), respectively. The gray regions define systematic uncertainties. Panel (b): flux variation in time of the solar disk component above 1 GeV compared with the SSN.  Panel (c): flux variation in time of the solar disk component above 1 GeV compared with the CR proton flux at 2.14--2.4~GV \citep{AMSprotons}.  Panel (d): flux variation in time of the solar disk component above 1~GeV compared with the Moon's flux above 56~MeV \citep{moon2016}. 
            \newline
            \label{fig1}}
        \end{minipage}
        \hspace{0.05\textwidth} % Horizontal spacing between figures
        % Right figure
        \begin{minipage}[b]{0.45\textwidth} % 45% of text width
            \centering
            \includegraphics[width=\textwidth]{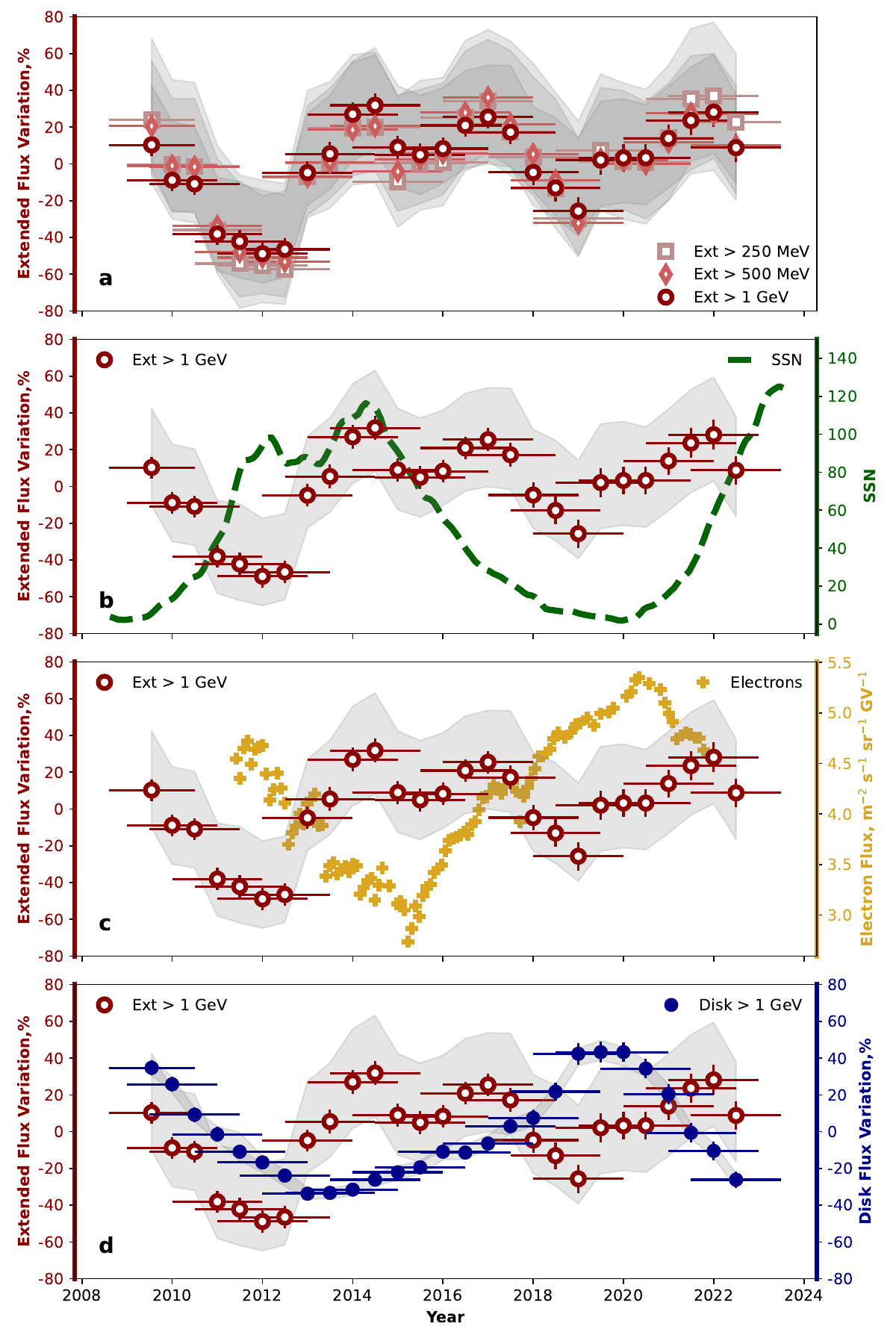} % Replace with your image
            \caption{Panel (a): Percentage of the flux variation of the solar extended component above 250~MeV, 500~MeV, and 1~GeV versus time. For normalization, the average flux values used are 30.00, 13.99, 5.95 ($\times$ 10$^{-8}$ cm$^{-2}~$s$^{-1}$), respectively. The gray region defines systematic uncertainties. Panel (b): flux variation of the solar extended component above 1 GeV compared with the SSN. Panel (c): flux variation of the solar extended component above 1~GeV compared with the CR electron flux at 1.00--1.71~GV \citep{AMSelectrons}. Panel (d): flux variation of the solar extended component above 1~GeV compared with the flux variation of the disk component above 1 GeV. 
            \newline
            \label{fig2}}
        \end{minipage}
   }
\end{figure*}

\subsection{Disk Component} 

Figure~\ref{fig1} shows details of the disk emission. Panel (a) shows the flux variation expressed as a fraction of the average flux of the disk component for energies greater than 250 MeV, 500 MeV, and 1 GeV over time. 
The regions with different shades of gray represent the systematic uncertainties. The variations in the three integral energy ranges are synchronous and the variation of the disk flux with the solar cycle is clearly demonstrated. 
For all three energy ranges, the disk flux varies by $\pm40\%$ with respect to the average value. 
Panel (b) shows the anticorrelation between the disk flux variation above 1~GeV and the SSN over time. This is consistent with expectations, as CR-induced emission is expected to vary over the solar cycle in anticorrelation with solar activity \citep[e.g.,][]{Thompson, Ng}.
Panel (c) shows correlation between the disk flux variation above 1~GeV and the temporal variation of the CR proton flux at $\sim$2 GeV as observed by AMS-02. 
Hence, the hadronic emission of the disk component around GeV energies is correctly identified. This also means that the modulation of CR protons at Earth is temporally coincident with the modulation at the Sun within the 2-year resolution of the gamma-ray fluxes.   
Panel (d) shows correlation between the solar disk flux above 1~GeV and the lunar flux \citep{moon2016}. The lunar flux variations throughout the period \citep{Loparco2024} show a similar trend. This corroborates the theory that both solar disk and lunar emission share a similar production mechanism. The comparisons in Panels (b), (c), (d) give Pearson and zDCF correlations (or anti-correlations) above 0.9\footnote{We find this value for the non-overlapping 2-year sub-samples, and also for all data points.} with no significant time lag.

\subsection{Spatially Extended Component} 
Figure~\ref{fig2} shows details of the extended component. Panel (a) shows the flux variation expressed as a fraction of the average flux of the solar extended component for energies greater than 250 MeV, 500 MeV, and 1 GeV  over time. 
The gray-shaded regions represent its systematic uncertainties.   
For the three integral energy ranges, the flux varies by +40\% and $-$60\% between its maximum and minimum values. In comparison to the disk emission, the systematic errors for the extended component are larger due to the uncertainties in the background estimation.

The extended flux variation above 1 GeV is compared with the sunspot number (panel (b) of the same figure) and with the temporal variation of the CR electron\footnote{The IC component also includes positrons, but their flux accounts for at most a few percent of the total.}
flux at 1 GeV as observed by AMS-02 (panel (c)). 
The plots show a complex relationship between the extended component flux variation and both the SSN and CR electron flux. Indeed, the correlation coefficients over the entire period are below 0.3, indicating no significant correlation.
Because the periods before 2011 lack CR electron measurements from AMS-02, to gain insights on this complex trend of the extended component, panel (d) shows a temporal comparison between the flux variations of the extended component and the disk. 
A visual comparison of panels (b), (c), and (d) appears to reveal periods consistent with expectations, as well as periods contrary to them. The correlation coefficients with the disk flux variation and the sunspot number, calculated only for the period from August 2008 to July 2012, show strong correlation and anti-correlation (above 0.9), respectively, in line with expectations.

%Comparisons with the disk variation and the sunspot number from August 2008 to July 2012 give a Pearson and zDCF correlation and anti-correlation respectively above 0.9, as we would expect for the full period} \citep[e.g.,][]{stellarics}. %and give no correlation (or anti-correlation) after 2013. 

\section{Discussion and Conclusions} \label{sec:floats}

We report for the first time the temporal variation of the flux of both distinct gamma-ray components of the solar emission over 15 years of \textit{Fermi}-LAT observations. 
In the following, we discuss the two solar emission components separately.

The flux variation in time of the disk anticorrelates with the sunspot number and correlates with cosmic-ray protons, confirming its emission mechanism.
It generally agrees with the results of \cite{Linden2022} obtained for a shorter period. Possible differences are attributable to different temporal binning and data selections. Interestingly, we find that the variation in time is independent of energy above 250~MeV, whereas the flux in the lowest energy range is expected to vary the most\footnote{Calculations in \citet{Mazziotta} around solar maximum might suggest weak energy independence, though this remains speculative due to limited coverage and the absence of calculations during solar minimum.} due to the stronger modulation of CRs at lower energies. The fact that the relative variation of fluxes does not depend on energy was already anticipated by \cite{Linden2022}. It can be an indication that the solar magnetic field plays an important role in the variability, as the intricate structure of the field at the Sun would lead to complex energy and time variations. 

The variation of the extended component has not been investigated before\footnote{After completing this work and the manuscript, we became aware of a related study by \cite{Linden2025}, which had just appeared on arXiv.}. 
We find temporal anticorrelation between the extended flux variation and the SSN from 2008 August to 2012 June, and correlation between the extended flux variation and the disk variation over the same period.
However, for 2013–2023, we no longer observe any correlation/anticorrelation, not even with the CR electron flux. Similarly to the disk, the variation is independent of energy.

The variations of the extended component come as a surprise and challenge current theoretical models and assumptions. 
We do not provide a definitive interpretation of these results; instead, we highlight related observational and modeling challenges to illustrate the complexity of the topic.
For example, in the outer heliosphere, it has long been accepted that  CR electrons are modulated similarly to protons \citep[e.g.,][]{Gleeson} and that their modulation anticorrelates with the SSN. 
However, several studies on heliospheric CR propagation beyond 1 AU have reported indications, based on measurements \cite[e.g.,][]{AMSmod2018} and models \cite[e.g.,][]{Potgieter2013,Potgieter2017,Tomassetti2017,Tomassetti2022,CL2019}, that CR modulation and propagation depend on the particle’s charge sign and the solar magnetic polarity.
For protons, \cite{Tomassetti2017} proposed a simple predictive model of solar modulation in the outer heliosphere that depends on the SSN and the tilt angle to explain the observed time lag between solar activity data and CR measurements. This time lag is found to depend on the solar polarity. 
Recently, in order to explain the modulation over time of the AMS-02 CR electrons, \cite{Aslam} calculated that particle drift in the outer heliosphere becomes negligible during the period when the polarity is not well defined and starts recovering just after the polarity reversal, but the mean free path keeps decreasing or remains unchanged for some period after the polarity reversal.
On the other hand, the propagation of CR electrons, in the inner heliosphere, i.e., within 1~AU of the Sun, has received comparatively little attention.
A recent study on cosmic-ray transport in the inner heliosphere \citep{POS2023} shows that electrons undergo multiple scattering due to turbulence and experience significant energy losses as they approach the Sun. Based on this, we expect that at higher energies the reduced electron flux near the Sun under stronger magnetic fields would decrease the IC emission, reinforcing the expected anticorrelation with solar activity.

While most of the current literature focuses on CR propagation in the outer heliosphere rather than in the inner heliosphere, our present results suggest that the details of CR electron transport and modulation in the inner heliosphere are even more complex than in the outer heliosphere. Hence, models of CR electron propagation that work beyond 1 AU may not be extrapolated down to the Sun. 
Modulation and propagation within 1 AU of the Sun have rarely been studied because of poor observational constraints. Our work can shed light on the propagation of CRs in the inner Galaxy through the study of the variation of the CR-induced emission from the Sun.

Curiously, at the end of 2012, the reversal of the Sun's polar magnetic field began, which resulted in the change of polarity in Cycle 24. 
In contemporary with the reversal, unexplained asymmetric emission from the Sun in the GeV range has recently been observed \citep{ArsioliOrlando}. Interestingly, the polarity reversal was unusual \citep{Janardhan, upton} with the south polarity flipping in 2013, while the north polarity being delayed to the end of 2014.

Our results also suggest, as do other studies \cite[e.g.][]{Tang, Linden}, a possible crucial role of the solar magnetic field, which remains unexplained. 
Possibly, CR electron transport and modulation models in the inner heliosphere may be constrained by observations of synchrotron radiation in X-rays produced by these electrons interacting with the solar magnetic field, as recently theorized by \citet{Orlando2023}. Other effects may also play an important role \citep[e.g.,][]{Puzzoni,Ng2024,MazziottaSun}.

Our results may even call into question whether the spatially extended emission is produced mainly by Galactic CR electrons. This may potentially include an unknown additional mechanism of gamma-ray production or an additional component of high-energy CR close to the Sun. 
Both hypotheses would need to be investigated further. \\

To conclude, the discovery reported in this work presents an additional challenge to our understanding of the Sun in gamma rays. This adds to the detection of the solar unexplained spectral dip \citep{Linden}, the observations of an unexpected asymmetry in GeV emission from the disk \citep{ArsioliOrlando}, and the detection of GeV and TeV solar emission at higher energies than expected \citep{Abdo2011, Linden, HAWC}.
A revision of current models and timely observations of CRs and the Sun during the current Cycle 25, as well as during the reversal in polarity, all will help in understanding the unexpected trend of the solar extended emission and its relation to the solar cycles.

\section{Acknowledgments} 
%We thank the anonymous referee for the useful comments.
The \textit{Fermi}-LAT Collaboration acknowledges generous ongoing support
from a number of agencies and institutes that have supported both the
development and the operation of the LAT as well as scientific data analysis.
These include the National Aeronautics and Space Administration and the
Department of Energy in the United States, the Commissariat \`a l'Energie Atomique
and the Centre National de la Recherche Scientifique / Institut National de Physique
Nucl\'eaire et de Physique des Particules in France, the Agenzia Spaziale Italiana
and the Istituto Nazionale di Fisica Nucleare in Italy, the Ministry of Education,
Culture, Sports, Science and Technology (MEXT), High Energy Accelerator Research
Organization (KEK) and Japan Aerospace Exploration Agency (JAXA) in Japan, and
the K.~A.~Wallenberg Foundation, the Swedish Research Council and the
Swedish National Space Board in Sweden.
Additional support for science analysis during the operations phase is gratefully
acknowledged from the Istituto Nazionale di Astrofisica in Italy and the Centre
National d'\'Etudes Spatiales in France. 
This work was performed in part under DOE
Contract DE-AC02-76SF00515.
Partial support from NASA grant No. 80NSSC20K1558 and  80NSSC22K0495 is acknowledged.

\end{document}